\documentclass[twocolumn,aps,pre]{revtex4}
\usepackage{graphicx,color}
\begin{document}
\title{\bf{Renormalization group analysis  for  an asymmetric 
simple exclusion process}}
\author{Sutapa Mukherji}
\affiliation{Protein Chemistry and Technology, Central Food Technological Research Institute, Mysore -570 020, Karnataka}
\date{\today}
\begin{abstract}
A perturbative renormalization group method is used  to obtain  
steady-state density profiles of a particle non-conserving asymmetric 
simple exclusion process.   This method allows us to obtain a globally valid 
solution for the density profile without the asymptotic matching  of bulk and boundary layer 
solutions. In addition, we  show  a nontrivial scaling of the boundary layer width with the 
system size close to specific  phase boundaries.
\end{abstract}
\maketitle
 Boundary layers are found to play an important role in  boundary 
induced phase transitions of  asymmetric 
simple exclusion processes (ASEPs). In ASEP,  particles 
hop on a finite one dimensional lattice with a bias in a 
preferred direction.  In addition, particles obey
 mutual exclusion principle that prohibits two particles from 
occupying the same lattice 
site \cite{liggett,derrida,schuetz}. 
In the steady state, the average particle distribution 
profile on the lattice depends completely on the particle 
injection rate, $\alpha$, at one end of 
the lattice  and withdrawal rate, 
$\beta$, at the other end.  Boundary layers appear in 
the particle distribution profile and, in general,  their   width, height, and   
location  change sharply with the injection and withdrawal 
 rates  \cite{smsmb}. These rates are the tuning 
parameters for the boundary induced phase transitions which 
can be characterized through the  shape and 
the properties of the density profile and, in particular,  its boundary layer.

In order to obtain the steady-state density profile 
 of the ASEP in the hydrodynamic limit, one has to solve    
a boundary value problem involving singular nonlinear 
differential equation(s)  for the particle 
density profile with boundary conditions given by the boundary rates \cite{smsmb,cole}. 
The singularity in the  equation  is due to the 
highest order derivative term  that 
appears along  with a  small multiplicative paramater, $\epsilon$, which 
is inversely proportional to the system size, $N$. 
Solutions of such equations 
have boundary layers which are narrow segments 
over which the solution varies rapidly 
in comparison with the remaining, slowly varying  (bulk) part  of the solution . 
Here by "boundary layer', we imply a  solution of the differential equation
under a specific limit. This  solution need not necessarily appear 
only near the system boundary.  In the same spirit, the rest of the density profile is 
referred as the bulk solution which also need 
not be confined to  the interior of the system.  
In particular, in ASEP, the bulk solution often satisfies one of the boundary conditions. 
 Methods of boundary layer analysis can be used to find a uniform 
approximation for the solution of the  boundary value problem. 
This is achieved  by first constructing 
 the asymptotic expansions for the solutions near the  
boundary-layer  and the bulk  regions  and then joining these solutions  
by  matching them in the appropriate limit. 
 A more general method to obtain the density profile including its 
 boundary layers involves obtaining 
the fixed points of the boundary layer differential equation \cite{smfixedpt}. Since 
the boundary layer saturates to the bulk solution, one can extract information 
about the bulk solution by finding out the 
boundary layer fixed points and their stability properties.

Although    boundary layer based methods
are useful in many cases,  
the scaling behavior  of the 
width of the boundary  layer with the system size 
 may not be fully uncovered through this method. 
Previous boundary-layer based work 
shows that the scaling of the boundary layer width follows 
 simple dimensional analysis. 
This, however, need not be the case always, since 
Monte-Carlo simulations on certain ASEP models show 
deviations from simple dimensional analysis \cite{parmeg}. 
Historically, renormalization group (RG) analysis                                
 appears to be one of the efficient tools 
that can be used in  situations
where simple dimensional considerations do not work \cite{amit}.
 The application of RG analysis to singularly perturbed 
 differential equations  shows that the presence 
of multiple scales in the problem comes out as a natural 
consequence of the analysis and no apriori knowledge  
is required for obtaining multiple scales \cite{oono}. 
The utility of the renormalisation group approach in solving various 
singular perturbation problems  has  been further elaborated later in \cite{mallet}.
In view of this, a natural question arises as whether RG  analysis can reveal  
the presence of multiple scales, if any,  in ASEP.

The main purpose of the paper is to present an RG analysis for the 
  ASEP described here and attempt to find out if 
there exists any hidden  scaling not seen  so far through   the 
boundary layer method. As we shall show below, this analysis 
indeed shows the possibility of a different scaling near certain 
phase boundaries of ASEP. Apart from this, 
RG has additional benefits over conventional 
boundary layer methods \cite{oono,goldenfeld}. 
It allows us to obtain  uniform, globally valid solutions without 
the need of any asymptotic matching as done  in  
boundary-layer methods. This simple ASEP model  demonstrates 
 how this goal is achieved. Similar to the boundary layer method, 
 RG analysis   starts 
with a naive perturbation expansion of the boundary layer 
equation. In both methods, the future steps rely on finding the 
the analytical solution 
of  the differential equation at $O(\epsilon^0)$ level; a task 
which may not be  feasible for complex ASEP models. 
However, in case of RG,  this difficulty of finding an exact 
 solution of the lowest order equation 
may be bypassed by guessing a naive solution. 
Although, this gives rise to certain 
ambiguities, RG is believed to be robust 
against such pitfalls \cite{goldenfeld}. 
In view of this, we hope that these  studies on the simplest 
ASEP model may further be extended to more complex processes.   
  
The specific model that we consider  here is totally asymmetric in 
the sense that particles move only in one direction along a 
lattice of $N$ sites provided 
the  target site is empty. In addition, we also consider 
particle non-conserving processes which include particle 
evaporation (adsorption) from (to)  an occupied (unoccupied)  
lattice site.  In the discrete picture, 
 the dynamics of the particle can be described in terms of  a variable $\tau_i$ 
 that denotes the particle occupancy of the $i$th site. This variable 
 can have values $\tau_i=1 \ {\rm or}\  0$ if the $i$th site is occupied or empty, respectively.  
The time evolution of the  variable $\tau_i$ can be expressed as
\begin{eqnarray}
\frac{d\tau_i}{dt}=\tau_{i-1}(1-\tau_i)-\tau_i(1-\tau_{i+1}) +
\omega_a(1-\tau_i)-\omega_d \tau_i, 
\end{eqnarray}
where the first two terms on the right hand side  of the equation 
arise from particle hopping to the neighboring site and the 
last two terms are the gain and loss terms due to particle 
adsorption and evaporation at  rates $\omega_a$ and $\omega_d$, 
respectively. We shall consider a simple continuum mean-field 
description which requires first a statstical averaging of the 
time evolution equation  with the approximation 
$\langle\tau_i\tau_j\rangle=\langle\tau_i\rangle\langle\tau_j\rangle$  
and then going over to a continuum limit that involves 
 $N\rightarrow \infty$, lattice spacing $a\rightarrow 0$ limits  with $Na$ 
remaining finite. For simplicity, we choose $Na=1$ in the following.

\begin{figure}[ht!]
  \centering
   \includegraphics[height=0.4\textwidth]{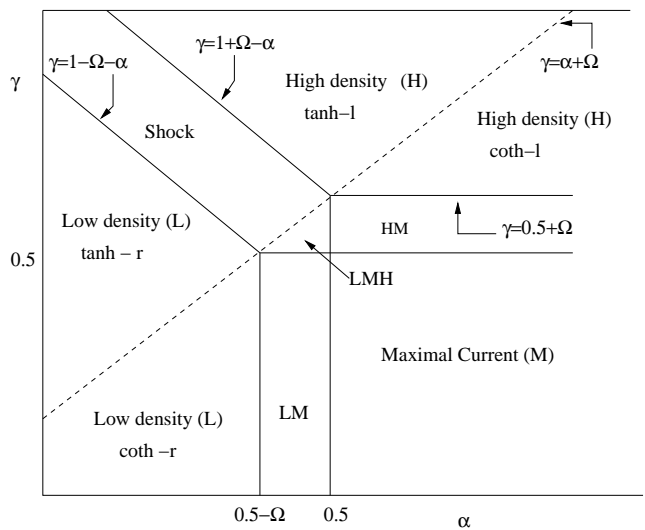}  
\caption{The phase diagram of the particle non-conserving  totally asymmetric simple exclusion 
process with $\omega_a=\omega_d$. HM and LM denote high-density-maximal-current  and low-density-maximal-current phases where  the density profile has a linear part (similar to high-density or low-density phases, see equation (\ref{outer})) 
 along with a  constant profile  (similar to the maximal current phase). In a similar way, LMH denotes 
 a phase where the density profile has parts similar to that of low-density (L), high-density (H) and maximal (M) current phases.   }
 \label{fig:phasediag}
\end{figure}

In the continuum limit, the steady state ($\frac{d\rho}{dt}=0$)  particle density, 
$\rho(x)$, at position $x$,
satisfies the differential equation 
\begin{eqnarray}
\epsilon\frac{d^2\rho}{dx^2}+(2\rho-1) \frac{d\rho}{d\tilde x}+
\Omega  (1-2\rho)=0. \label{hydro2},
\end{eqnarray}
where $\epsilon=1/(2N)$. 
First two terms in this equation can be expressed as the gradient 
of the current $-\frac{\partial J}{\partial x}$,  with 
$J=-\epsilon\frac{\partial\rho}{\partial x}+\rho(1-\rho)$.  The third term 
appears due to particle  loss and gain at equal rates ($\omega_a=\omega_d$)
with $\Omega=\omega_d N$. 
  In order to obtain the 
steady-state density profile, one needs to solve equation  (\ref{hydro2}) 
along with the boundary conditions $\rho(x=0)=\alpha$ and 
$\rho(x=1)=\gamma=1-\beta$ at the two ends of the lattice.  The 
particle conserving model has been solved exactly \cite{derrida, schuetz} 
and it appears that the phase diagram broadly consists of three phases 
with average  bulk particle density remaining constant  in each phase. 
For $\alpha<1/2$ and 
$\beta>\alpha$, the system is in a low density phase where the 
average particle density in the bulk of the profile 
is constant at a value $\rho=\alpha (<1/2)$. The bulk density profile  
satisfies the boundary condition 
at $x=0$ and a boundary layer appearing near $x=1$ 
satisfies the boundary condition at $x=1$. 
 For $\beta<1/2$ and $\alpha>\beta$, the system is in a high density phase 
 where the average bulk 
 density is $\rho=1-\beta>1/2$. Since in  this case, the constant bulk profile satisfies the   
 boundary condition at $x=1$, the boundary condition at $x=0$ 
  is satisfied by a boundary layer.  For $\alpha,\beta>1/2$, 
 the average particle density is constant at a value $\rho=1/2$  with boundary layers 
 appearing at both ends of the system.  Since the particle current has the maximum value in this 
 phase, this phase is known as the maximum current phase.   In the particle 
non conserving model, the bulk density is not constant but there exist 
similar low-density ( $\rho<1/2$ in the bulk), high-density ($\rho>1/2$ in the bulk) 
and maximum current ($\rho={\rm constant}=1/2$) phases in addition to 
several other phases where the density profile has constant, 
maximum-current parts along with  nonconstant parts similar to 
those in high or low-density phases (see figure \ref{fig:phasediag}) \cite{parmeg,evans,smsmb}. 
 As we shall show below, the approach to the HM or LM phase 
from the high-density or low-density phases respectively 
 is  special since, very close to the   phase boundary,
the boundary-layer width  is  expected to scale with the system 
size as $ N^{-1/2}$ unlike its usual scaling as $N^{-1}$.

The boundary layer analysis for this totally asymmetric simple exclusion process 
 proceeds as follows. 
For very small $\epsilon$, one may ignore the second-order derivative term in
(\ref{hydro2}) and obtain a solution of the corresponding first order equation as  
\begin{eqnarray}
\rho_o(x)=\Omega x+c\label{outer}
\end{eqnarray} 
This, being the zeroth order solution of the original equation, is 
expected to describe most  of the density profile.
Since in  the boundary layer language, this solution 
is known as the outer solution, we  have denoted this solution as $\rho_o(x)$.  
Clearly, a solution of a first order equation 
cannot satisfy two boundary 
conditions and hence, the presence of a boundary layer becomes necessary. 
A boundary layer solution for such a second order differential equation is subjected to  two 
conditions.  For example, in the low or high density phase, the boundary layer 
satisfies the boundary condition at one end 
and merges to the bulk solution (outer solution)  in the other limit. In order to satisfy 
two conditions, the second 
derivative term of  (\ref{hydro2})
becomes necessary for the description of the boundary layer. 
 A dominant balance argument shows that for a 
boundary layer near $x\sim O(0)$, one requires a rescaling of $x$ 
as $\tilde x=\frac{x}{\epsilon}$.   
Considering the fact that there can be various possible locations 
of the boundary layer, for example,  near $x=0$ or  $x=1$ boundary 
or in the interior of the 
lattice, we reexpress equation (\ref{hydro2})  in terms of a scaled 
variable $\tilde x=\frac{(x-x_0)}{\epsilon}$, where $x_0$ indicates the location of the 
boundary layer.  For example, for  a  boundary layer located near $x=1$ boundary, we have $x_0=1$.  
The differential  equation in terms of $\tilde x$ is 
\begin{eqnarray} 
\frac{d^2\phi}{d\tilde x^2}+\phi \frac{d\phi}{d\tilde x}-
2\Omega \epsilon \phi=0, \label{bleqn1}
\end{eqnarray}
where $\phi=2\rho-1$. Ignoring the 
last term in the $\epsilon\rightarrow 0$ limit, we have the 
zeroth order boundary layer equation 
\begin{eqnarray}
\frac{d^2\phi}{d\tilde x^2}+\phi \frac{d\phi}{d\tilde x}=0, \label{phinner2}
\end{eqnarray}
which upon one integration  appears as 
\begin{eqnarray}
\frac{d\phi}{d\tilde x}+\frac{\phi^2}{2}=c_1,\label{phinner}
\end{eqnarray}
where $c_1$ is the integration constant. 
Since, the boundary layer is expected to saturate  to the bulk solution (outer solution) in the 
appropriate limit of $\tilde x$, we consider $c_1=\frac{\phi_b^2}{2}$, where $\phi_b=2\rho_b-1$ with 
 $\rho_b$ as the corresponding bulk density at the saturation edge.  
 Since the  solution  of  the boundary layer equation (\ref{phinner}) is usually called 
 the inner solution, we denote this solution as  $\phi_{\rm in}$.  The explicit solutions
  of (\ref{phinner}) are 
\begin{eqnarray}
\phi_{\rm in}=\phi_b \tanh[\frac{\phi_b}{2}(\tilde x+c_2)] \ \ {\rm and} \label{tanhbl1}\\
\phi_{\rm in}=\phi_b \coth[\frac{\phi_b}{2}(\tilde x+c_2)],\label{cothbl2}
\end{eqnarray}  
where $c_2$ is the second integration constant. 
In terms of $\rho$, the boundary layer or inner  solutions are 
\begin{eqnarray}
\rho_{\rm in}=\frac{1}{2}+(1/2-\rho_b) \tanh[\frac{(1-2\rho_b)}{2} (\tilde x+c_2)] \ \ \ {\rm and} \label{tanhbl}\\
\rho_{\rm in}=\frac{1}{2}+(1/2-\rho_b) \coth[\frac{(1-2\rho_b)}{2} (\tilde x+c_2)].  \label{cothbl}
\end{eqnarray}
The saturation of $\tanh$ and $\coth$ functions for large values of their arguments is responsible 
for saturation of the boundary layers to the bulk. 
In order to understand how various conditions  are satisfied by the boundary layer, let us consider the 
density profile of the 
low-density phase as an example. The boundary layer located near $x=1$ boundary satisfies 
the boundary condition at $x=1$  and saturates to  the bulk solution in the 
$\tilde x\rightarrow -\infty$ ($x< 1$) limit.  Such a boundary layer can be consistently described 
by equation  (\ref{tanhbl}) since $\rho_{\rm in}$  approaches $\rho_b$ in the $\tilde x\rightarrow -\infty$ 
 limit. The matching (saturation) with the outer solution in equation 
 (\ref{outer}) demands $\rho_o(x=1)=\rho_b=\rho_{\rm in}(\tilde x\rightarrow -\infty)$.

The RG analysis starts with the perturbative expansion for the boundary 
layer solution. The  renormalized perturbative theory, however,   gives the 
uniform globally valid  solution for the entire density profile. In order to find the 
solution of the boundary layer equation (\ref{bleqn1})  perturbatively, we assume 
a  naive expansion of the  solution as $\phi(\tilde x)=\phi_0(\tilde x)+
\epsilon \phi_1(\tilde x)+\epsilon^2\phi_2(\tilde x)+....$. While at $O(\epsilon^0)$, $\phi_0$ 
satisfies the same equation as (\ref{phinner}), at $O(\epsilon)$, we have  
\begin{eqnarray}
 \frac{d^2\phi_1}{d\tilde x^2}+\frac{d(\phi_0\phi_1)}{d\tilde x}-
2\Omega \phi_0=0.\label{order1}
\end{eqnarray}
We rewrite the solution of the $O(\epsilon^0)$ equation as  
\begin{eqnarray}
&& \phi_0=\sqrt{c} \coth[\frac{\sqrt {c}}{2}(\tilde x+k)] , \\  
&& \phi_0=\sqrt{c} \tanh[\frac{\sqrt{c}}{2}(\tilde x+k)],\label{solnnoncons}
\end{eqnarray}
where $c$ and $k$ are the integration constant with  $c={\phi_b^2}$. 
In order to find $\phi_1$, we  use the $\tanh$ solution for $\phi_0$ in (\ref{order1}). 
The solution for $\phi_1$ is 
\begin{eqnarray}
&&\phi_1= (2 p)^{-1} \text{sech}[\frac{1}{2} p(k+\tilde x)]^2
{\Bigg[}2 p \Omega(k+\tilde x)-\nonumber\\
&&\Omega p^2(k+\tilde x)^2+
 (k+\tilde x) p \overline{c}+
2 p \overline{k}-\nonumber\\ && 4 \Omega p (k+\tilde x) {\Big\{}\log[1+e^{-p(k+\tilde x)}]+
\nonumber\\ && \log{\Big[}\text{cosh}\{\frac{p}{2}(k+\tilde x)\}{\Big]}{\Big\}}+
 4 \Omega\ \text{Li}_2{\Big(}\exp[-p(k+\tilde x)]{\Big)}+\nonumber\\ && 
{\Big(}-2\Omega +\overline{c}+4\Omega 
\log{\Big[}\text{cosh}\{\frac{1}{2}p(k+\tilde x)\}{\Big]}{\Big)}\times\nonumber\\ &&\text{sinh}[p(k+\tilde x)]
{\Bigg]},\label{phi1fullform}
\end{eqnarray}
where $p=\sqrt{c}$,  $\overline{c}$ and $\overline{k}$ are constants of integration and $\text{Li}_n(y)$ is 
the Polylogarithm function of order $n$ and argument $y$.  In the boundary layer regime, 
the Polylogarithm function is expected to have real values. For example, 
 for the $\tanh$ boundary layer  present near $x=1$  in the low density phase, the argument of the 
 exponential function in the Polylogarithm is always large negative 
 ($\tilde x \rightarrow -\infty$ and $p=\phi_b=2\rho_b-1<0$). This makes the argument of Polylogarithm 
 negligibly small. 
The naive  perturbation done here breaks down due to the  term 
$4\Omega(2p)^{-1} {\rm sech}[\frac{1}{2}p(k+\tilde x)]^2\sinh[p(k+\tilde x)] \log{\big[}\cosh\{\frac{1}{2}p(k+\tilde x)\}{\big]}$ 
that diverges as $\tilde x\rightarrow \infty$. This term, when multiplied with $\epsilon$, 
becomes comparable to the zeroth order solution since  
$\epsilon \tilde x$  becomes $O(1)$ in the 
$\tilde x \rightarrow \pm \infty$ limit. The appearance of this 
divergence in the naive perturbation theory 
is the manifestation of the singularity in the problem. 

The perturbative solution upto $O(\epsilon)$ can be expressed as 
\begin{eqnarray}
&&\phi(\tilde x)= p\ \text{tanh}[\frac{p}{2}(k+\tilde x)]+
\epsilon {\cal R}(\tilde x)+\nonumber\\ 
&&\epsilon \frac{4\Omega}{p} 
\log[\text{cosh}\{\frac{1}{2}p(k+x)\}] 
\text{tanh}[p(k+x)/2],\label{phi1d}
\end{eqnarray}   
where the secular term is written clearly and all the 
 regular terms are represented together by ${\cal R}$.
Considering the  the $\tilde x\rightarrow \infty$ limit, in which 
$\log[\cosh[\frac{1}{2}p(k+\tilde x)]\sim \frac{1}{2}p(k+\tilde x)$, 
we  separate  the divergence in the perturbation series as 
\begin{eqnarray}
&& \phi(\tilde x)=p\ \text{tanh}[\frac{p}{2}(k+\tilde x)]+
\epsilon {\cal R}(\tilde x)+ \nonumber\\ 
&& \epsilon 2\Omega (k+\tilde x-\mu)\tanh[\frac{p}{2}(k+\tilde x)]+\nonumber\\
&& \epsilon 2\Omega \mu \tanh[\frac{p}{2}(k+\tilde x)],
\label{phi1d1} 
\end{eqnarray}
where  $\mu$ is an arbitrary length scale chosen in such a way that the 
second term is now no more diverging and the entire divergence is 
contained in the last term. The divergence is absorbed by introducing 
a renormalized constant, $p_r$, as $p=p_r(\mu)+Z_1(\mu)$
where $Z_1(\mu)=a_1 \epsilon+a_2 \epsilon^2...$.
The divergence in (\ref{phi1d1}) can be absorbed 
if  $a_1=-2\Omega \mu$.
The renormalized perturbation series for $\phi$ is now 
\begin{eqnarray}
&&\phi=p_r(\mu) \text{tanh}[\frac{p_r(\mu)}{2}(k+\tilde x)]+
\epsilon\ 2\Omega (k+\tilde x-\mu)\times\nonumber\\ 
&&\text{tanh}[\frac{p_r(\mu)}{2}(k+\tilde x)]\nonumber\\
&&-\epsilon\ \Omega \mu p_r(\mu)(k+\tilde x) 
\text{sech}[\frac{p_r(\mu)}{2}(k+\tilde x)]^2
+\epsilon {\cal R}(\tilde x),\label{phirg}
\end{eqnarray} 
where the last term arises from the expansion of the $\tanh$ term in small $\epsilon$.
The renormalized theory cannot depend on an arbitrary length scale 
$\mu$. The condition, $\frac{\partial\phi}{\partial \mu}=0$,  leads to the
following equation 
\begin{eqnarray}
&&\frac{\partial\phi}{\partial \mu}=\frac{d p_r}{d\mu}\tanh[\frac{p_r}{2}(k+\tilde x)]+
p_r \frac{dp_r}{d\mu}\frac{(k+\tilde x)}{2} \times\nonumber\\ 
&& {\rm sech}^2[\frac{p_r}{2}(k+\tilde x)]-{\epsilon}\Omega p_r (k+\tilde x) {\rm sech}^2[\frac{p_r}{2}(k+\tilde x)] -\nonumber\\
&& \epsilon \Omega\mu(k+\tilde x) {\rm sech}^2[\frac{p_r}{2}(k+\tilde x)] \frac{dp_r}{d\mu}+\epsilon \Omega\mu (k+\tilde x)^2\times\nonumber\\ 
&&{\rm sech}^2[\frac{p_r}{2}(k+\tilde x)]\tanh [\frac{p_r}{2}(k+\tilde x)] \frac{dp_r}{d\mu}\nonumber\\
&&-2\Omega\epsilon \tanh [\frac{p_r}{2}(k+\tilde x)]+2\Omega \epsilon(k+\tilde x-\mu) \frac{(k+\tilde x)}{2}\times\nonumber\\
&&{\rm sech}^2[\frac{p_r}{2}(k+\tilde x)]\frac{dp_r}{d\mu}=0
\end{eqnarray}
Up to $O(\epsilon)$, this leads to an RG equation 
\begin{eqnarray}
\frac{dp_r}{d\mu}=\epsilon (2\Omega).
\end{eqnarray}
Substituting the solution for, $p_r(\mu)$,  as
$p_r(\mu)\sim 2 \Omega \mu\epsilon+C_p$ in (\ref{phirg}), and setting 
$\mu=\tilde x$, we have the following  renormalized profile 
\begin{eqnarray}
&& \phi(x)=(C_p+2\Omega \epsilon \tilde x) \text{tanh}[(C_p+2\Omega \epsilon \tilde x)
(k+{\tilde x})/{2}]+\nonumber\\
&& \epsilon {\Big [}2\Omega\ k
\ \text{tanh}[(C_p+2\Omega\epsilon  \tilde x)(k+\tilde x)/2]-\nonumber\\
&& {\tilde x}(k+{\tilde x}) \Omega C_p
\text{sech}^2[(C_p+2\Omega \epsilon {\tilde x})(k+\tilde x)/2]{\Big]}\nonumber\\
&& +\epsilon {\cal R}(\tilde x), \label{renormphix} 
\end{eqnarray}
where $C_p$ is the integration constant. 
This solution is the globally valid solution that has a boundary layer 
of  $\tanh$ kind near $\tilde x\sim 0$ and a linear profile as in 
(\ref{outer}) for large $\tilde x$ limit ($+\infty$ or $-\infty$)
 where $\tanh$ saturates. 
 The first term in (\ref{renormphix}) approximates to a 
 linear solution  for $\tilde x\rightarrow \pm \infty$ limit. For small $\tilde x$, i.e. 
 near the boundary, the first term leads to a $\tanh$ type boundary layer (neglecting the 
 $\tilde x$ dependent term in the prefactor of the $\tanh$ term). 
 The second term, due to its prefactor, $\epsilon$  has a negligible 
 contribution for large $\tilde x$ as well as $\tilde x\approx 0$. In a 
 similar way,  the third term also has negligible contribution in the 
 $\tilde x\rightarrow \pm \infty$ (due to the $\rm{sech}$ term) and for
 $\tilde x\approx 0$.

In the high-density phase,  for certain parameter 
values, the density profile  has a  $\tanh$ type boundary layer 
near $x=0$ boundary. In the $\tilde x\rightarrow \infty$ limit, 
the boundary layer merges to 
a  linear profile satisfying the boundary condition at $x=1$. 
The integration constants of  (\ref{renormphix})  determined 
using  the boundary conditions
$\phi(\tilde x=0)=2\alpha-1$ and $\phi(x=1)=2\gamma-1$ are 
found as 
\begin{eqnarray}
C_p {k}= \ln\frac{\gamma+\alpha-\Omega-1}{\gamma-\alpha-\Omega} \  {\rm with} \ 
C_p=2\gamma-2\Omega-1.
\end{eqnarray}
In  the low density phase, the linear part of the density profile satisfies the 
boundary condition at  $x=0$ and the $\tanh$ type boundary layer satisfies the boundary 
condition at $x=1$. Thus,  imposing the condition $\phi(\tilde x=0)=2\gamma-1$ and 
$\phi(x=0)=2\alpha-1$, we have 
\begin{eqnarray}
C_p k=\ln\frac{\alpha+\Omega+\gamma-1}{\alpha+\Omega-\gamma} \  {\rm with} \ 
C_p=2\Omega+2\alpha-1.
\end{eqnarray}

The renormalized density profile with a $\coth$ type zeroth order solution 
can be derived in an identical way.  At the lowest order, the density 
profile has the form 
\begin{eqnarray}
\phi=(C_p+2\Omega \epsilon \tilde x) \text{coth}[(C_p+2\Omega \epsilon \tilde x)(k+{\tilde x})/{2}].
\end{eqnarray}
In the high density phase, for certain range of parameter values, one may 
have a  $\coth$ boundary layer satisfying the boundary condition at $x=0$. 
In this case, the constants in the renormalised profile are 
\begin{eqnarray}
C_p=2\gamma-2\Omega-1,\ \  {C_p k}=\ln\frac{\gamma-\Omega+\alpha-1}{\alpha-\gamma+\Omega}.
\end{eqnarray}
 Density profiles of this 
shape is observed in the high-density phase, on the  right side of the 
line $\gamma=\alpha+\Omega$ (see figure \ref{fig:phasediag}). 
Further,  as one proceeds near 
its phase boundary, $\gamma=1/2+\Omega$,
 demarcating the high-density and high-density-maximum current (HM) phase, the new scaling 
of the boundary-layer  width with the system size $N$ as $N^{-1/2}$, takes over.   Similar scaling behavior of the boundary layer width  is seen 
 as one approaches the phase boundary $\alpha=1/2-\Omega$ 
  between the low-density and the LM phase from the low density side. 
  
In conclusion, we have  presented a 
perturbative renormaization group analysis for a particle non conserving 
totally asymmetric simple exclusion process. 
 Such systems with open boundaries are found to exhibit boundary driven phase transitions in the steady state. The shape of the average particle density profiles 
 in various phases depends crucially on the boundary parameters which, in this case, are the 
 particle 
 withdrawal and injection rates at the boundaries.  In various phases,  particle density profiles,  in  general,  have rapidly varying, narrow boundary layer parts and slowly varying  bulk  parts.    
 A complete solution for the density profile can be obtained by using the method of boundary layer analysis which   provides a systematic tool to solve the   singular differential equation 
which, in the present case,   describes the steady state particle density profile.  
The full density profile  is obtained by finding  the 
boundary layer and the bulk solutions of the singular differential equation and 
 matching these two  solutions  in appropriate limits. 
Here we present a renormalisation group analysis 
 of  the singular differential equation  and show that 
 through this method one can arrive at a global solution of the differential equation without 
 any  asymptotic matching of the boundary layer and the bulk solutions. 
 The globally valid, general  solution, thus  obtained,  
contains in itself  both the boundary layer   and the   bulk parts which can be 
individually identified by considering different limits of the general solution. 
Further, it is shown that   as the system approaches the low-density-maximal current phase or the
   high-density-maximal current phase from the low-density or high-density sides, respectively, 
   the boundary layer width scales as $N^{-1/2}$ rather than $N^{-1}$.

{\bf Acknowledgement} I thank S. M. Bhattacharjee  for useful discussions.
Financial support from the   Department of Science
and Technology,  India   and warm hospitality of ICTP (Italy),
 where the work was initiated, are  gratefully acknowledged.


\begin{thebibliography}{99}
\bibitem{liggett} T. Liggett, {\it Interacting Particle Systems:
Contact, Voter and Exclusion Processes} (Springer-Verlag, Berlin, 1999).
\bibitem{derrida} B. Derrida, M. R. Evans, V. Hakim and  V. Pasquier, J. Phys. A {\bf 26}, 1493 (1993).
\bibitem{schuetz} G. Schuetz and E. Domany, J. Stat. Phys. {\bf 72}, 277 (1993).
\bibitem{smsmb} S. Mukherji and S. M. Bhattacharjee, J. Phys. A
{\bf 38}, L285 (2005); S. Mukherji and V. Mishra,
Phys. Rev. E {\bf 74}, 01116 (2006).
\bibitem{cole}J. D. Cole, {\it Perturbation Methods in Applied Math-
ematics}  (Blasidal Publishing company, Walthum, MA: 1968).
\bibitem{smfixedpt} S. Mukherji, Phys. Rev. E {\bf 79}, 041140 (2009).
\bibitem{parmeg} A. Parmeggiani, T. Franosch and E. Frey,   Phys. Rev. E {\bf 70} 046101 (2004);
see also \cite{evans}.
\bibitem{amit} D. J. Amit, {\it Field Theory, the renormalisation Group and Critical Phenomena} 
(World Scientific, Singapore, 1984).
\bibitem{oono} Lin-Yuan Chen, Nigel Goldenfeld and Y. Oono, Phys. Rev. E {\bf 54}, 376 (1996)
\bibitem{mallet} R. E. O'Mallet, Jr. and E. Kirkinis,
 Studies in applied mathematics, Vol 124 (2010).
 \bibitem{goldenfeld}John Veysey II and Nigel Goldenfeld,  Rev.  Mod. Phys.{\bf 79}, 883 (2007). 
 \bibitem{evans} M. R.  Evans,  R. Juhasz  and L. Santen   Phys. Rev. E {\bf 68} 026117 (2003).
\end{thebibliography}
\end{document}